\documentclass[sigconf]{acmart}

\usepackage{amsmath,amsthm}
\makeatletter
\@ifundefined{Bbbk}{}{}
\@ifundefined{openbox}{}{}
\makeatother
\usepackage{amssymb}

\usepackage{booktabs}
\usepackage{enumitem}
\usepackage{graphicx}
\usepackage{xcolor}
\usepackage{listings}
\usepackage{float}
\usepackage{algorithm}
\usepackage[noend]{algpseudocode}
\usepackage{tikz}
\usetikzlibrary{arrows.meta,positioning,fit,calc,shapes,decorations.pathmorphing}

\newtheorem{problem}{Problem}

\usepackage{amsmath,amssymb}
\usepackage{enumitem} 
\usepackage{wrapfig}   
\begin{document}

\title{Hardware Software Optimizations for Fast Model Recovery on Reconfigurable Architectures}


\author{Bin Xu}
\email{binxu4@asu.edu}
\orcid{0000-0001-6529-1644}
\affiliation{%
  \institution{IMPACT lab, Arizona State University}
  \city{Tempe}
  \state{Arizona}
  \country{USA}
}

\author{Ayan Banerjee}
\email{Ayan.Banerjee@asu.edu}
\orcid{0000-0000-0000-0000}
\affiliation{%
  \institution{IMPACT lab, Arizona State University}
  \city{Tempe}
  \state{Arizona}
  \country{USA}
}

\author{Sandeep K.S. Gupta}
\email{Sandeep.Gupta@asu.edu}
\orcid{0000-0002-6108-5584}
\affiliation{%
  \institution{IMPACT lab, Arizona State University}
  \city{Tempe}
  \state{Arizona}
  \country{USA}
}

\begin{abstract}
Model Recovery (MR) is a core primitive for physical AI and real-time digital twins, but GPUs often execute MR inefficiently due to iterative dependencies, kernel-launch overheads, under-utilized memory bandwidth, and high data-movement latency. We present \textbf{MERINDA}, an FPGA-accelerated MR framework that restructures computation as a streaming dataflow pipeline. MERINDA exploits on-chip locality through BRAM tiling, fixed-point kernels, and the concurrent use of LUT fabric and carry-chain adders to expose fine-grained spatial parallelism while minimizing off-chip traffic. This hardware-aware formulation removes synchronization bottlenecks and sustains high throughput across MR’s iterative updates. On representative MR workloads, MERINDA delivers up to \textbf{6.3$\times$} fewer cycles than FPGA-based LTC baseline, enabling real-time performance for time-critical physical systems.
\end{abstract}

\keywords{Model Recovery, GRU, FPGA Acceleration}


\maketitle

\section{Introduction}
One of the central advances in the AI revolution is physical AI, where computational agents interact with—and learn from—physical systems for control and continual adaptation~\cite{radanliev2021artificial}. In edge AI settings, these agents must operate under tight latency, power, and privacy constraints, making physics-guided predictive inference especially valuable. A key enabler is model recovery (MR): extracting first-principles–guided dynamical equations from real-world data so the learned model serves as a digital twin (DT). Unlike purely data-driven models, the recovered DT supports online monitoring of safety, integrity, and unknown errors, while feedback from forward simulation can be used to adjust system responses in real time~\cite{tao2018digital,xu2025accelerated}.

\begin{figure*}[!t]
  \centering
  \includegraphics[width=\textwidth]{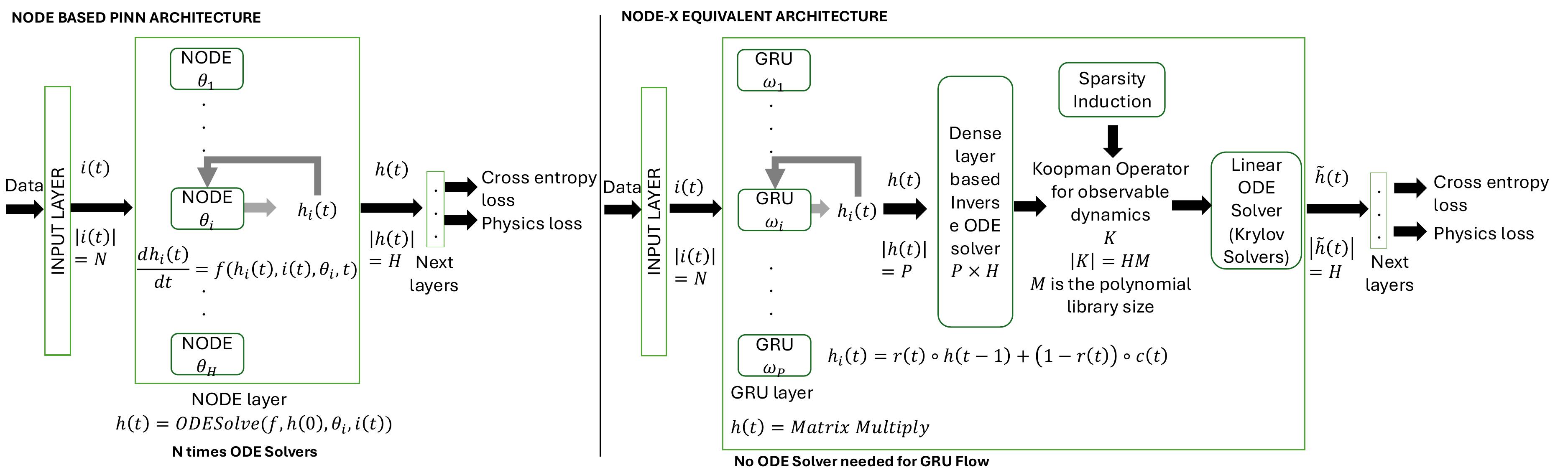}
  \caption{Equivalent GRU-based neural network (NN) architecture for MR.}
  \label{fig:NODEEq}
\end{figure*}

Traditionally, \emph{Model Recovery} (MR) relies on \emph{Neural Ordinary Differential Equations} (NODEs), which underpin continuous-depth residual networks and continuous normalizing flows~\citep{chen2018neural}. This paradigm extends to \emph{Liquid Time-Constant} (LTC) networks, which attain state-of-the-art sequence modeling by modulating input-driven nonlinear dynamical systems~\citep{liquid-time-constant-networks}. Despite their expressivity, these models depend on iterative ODE solvers during both training and inference, incurring substantial compute, latency, and energy costs. As illustrated in Fig.~\ref{fig:NODEEq}(left panel), a single forward pass through a NODE layer typically requires $N$ \emph{function evaluations} to advance the state. Because each sub-step depends on the previous one, these computations are inherently sequential and difficult to parallelize across steps, limiting throughput and amplifying launch/memory overheads on conventional accelerators.

\begin{table}[ht]
\centering
\scriptsize           
\setlength{\tabcolsep}{4pt} 
\renewcommand{\arraystretch}{0.9} 
\caption{Overall Forward Pass.}
\label{tab:forward-pass}
\begin{tabular}{lrr}
\toprule
\textbf{Operation} & \textbf{Time (ms)} & \textbf{Share (\%)} \\
\midrule
Sensory Processing   & 0.012997 & 12.3\% \\
ODE Solver (6 steps) & 0.093045 & 87.7\% \\
\midrule
\textbf{Total Forward Pass} & \textbf{0.106042} & \textbf{100.0\%} \\
\bottomrule
\end{tabular}
\end{table}

\begin{table}[ht]
\centering
\scriptsize
\setlength{\tabcolsep}{4pt}
\renewcommand{\arraystretch}{0.9}
\caption{ODE Step Breakdown (per step).}
\label{tab:ode-step}
\begin{tabular}{lrr}
\toprule
\textbf{Operation} & \textbf{Time (ms)} & \textbf{Share (\%)} \\
\midrule
Recurrent Sigmoid     & 0.007236 & 46.7\% \\
Weight Activation     & 0.000370 & 2.4\%  \\
Reversal Activation   & 0.000393 & 2.5\%  \\
Sum Operations        & 0.005341 & 34.4\% \\
Euler Update          & 0.002169 & 14.0\% \\
\midrule
\textbf{Single ODE Step Total} & \textbf{0.015507} & \textbf{100.0\%} \\
\bottomrule
\end{tabular}
\end{table}

\noindent\textbf{High-Level Optimization.}
Prior accelerators for ODENet~\cite{Wtanabe} and NODE layers~\cite{CaiFPGA} typically assume a \emph{fixed} solver depth and \emph{static} coefficients, which conflicts with data-driven MR frameworks (PINN+SR, PiNODE, EMILY) whose solver depth, step size, and parameters adapt to the input. Stand-alone ODE-solver engines~\cite{stamoulias2017high,ebrahimi2017evaluation} make similar assumptions, limiting their utility when coefficients vary online or when low-latency, streaming execution is required at the edge.

Profiling highlights the \emph{ODE solver} as the dominant cost: in Table~\ref{tab:forward-pass}, it accounts for 87.7\% of forward-pass latency (pre-processing 12.3\%) for a total of \(0.106\,\mathrm{ms}\). The per-step breakdown in Table~\ref{tab:ode-step} identifies \emph{Recurrent Sigmoid} (46.7\%) and \emph{Sum Operations} (34.4\%) as the main hotspots, followed by the Euler update (14.0\%), while weight and reversal activations contribute \(<5\%\). These findings show the bottleneck of computing N times ODE-Solvers during forward pass and backpropagation and motivate our substitution of the NODE layer with an \emph{equivalent, FPGA-friendly block} (Fig.~\ref{fig:NODEEq}, right): a GRU followed by a lightweight dense non-linearity and a single-step ODE solver. 
Concretely, the block computes a gated increment based on the current state and input, and then updates the state by adding this increment scaled by the step size, thereby preserving the original NODE mapping for both training and inference while eliminating multi-step solver overhead.

\noindent\textbf{Low-Level Optimization} 
To our knowledge, MR has not been systematically evaluated on FPGA platforms. This paper closes that gap by implementing and characterizing MR on edge-class FPGAs. 
Moving MR learning onto FPGAs can reduce data-transfer volume, cut end-to-end latency, and lower both storage and energy overheads by exploiting on-chip locality and streaming execution. 
However, this shift introduces its own challenges: unlike cloud GPUs, FPGAs provide more limited raw compute throughput, have far smaller on-chip and external memory capacity, and lack the mature software ecosystem of GPU platforms. 
These disparities raise fundamental questions about whether MR workloads—traditionally designed for GPU-friendly dense linear algebra—can meet real-time latency, energy, and memory constraints when deployed at the edge on FPGA-based systems.


FPGA toolchains (synthesis, mapping, placement, and routing) typically compile high-level designs into hardware by automatically mapping them onto computational and memory resources such as LUTs, DSPs, and adders. 
However, this automatic routing and mapping often yields suboptimal performance for streaming workloads such as GRU, due to inefficient memory access patterns and long routing paths\cite{lau2024rapidstream}. 
To address this, we introduce a \textbf{custom data-mapping strategy} that aligns memory layout with compute parallelism in the GRU architecture, ensuring each unrolled DSP MAC lane is supplied with operands every cycle. 
Furthermore, to sustain high throughput across pipeline stages, we apply \textbf{BRAM optimizations}---including partitioning, banking, and reshaping---to maximize on-chip bandwidth and reduce contention, thereby improving data movement efficiency for concurrent \texttt{DATAFLOW} execution.


Building on neural-flow theory~\cite{bilovs2021neural}, we present \textbf{MERINDA} (Model Recovery in Dynamic Architecture): a hardware-conscious reformulation of NODE layers (as used in EMILY, PiNODE, and PINN+SR) expressly tailored for FPGA acceleration. MERINDA replaces the iterative ODE solver with an invertible GRU-based block plus a dense nonlinearity, recasting MR as a streaming dataflow pipeline that maps naturally to FPGA fabrics. The design exploits (1) \textbf{on-chip locality} via BRAM/URAM tiling and banking to minimize off-chip traffic; (2) \textbf{spatial parallelism} by unrolling gate computations across DSP48 MAC lanes and evaluating sigmoid/$\tanh$ in LUT; (3) \textbf{fixed-point arithmetic} (8--16\,bits activations; 12--16\,bits weights/accumulators) to reduce area and power while preserving fidelity; and (4) \textbf{stage-level concurrency} using \texttt{DATAFLOW} with BRAM FIFOs so stages overlap and sustain $\mathrm{II}\!\approx\!1$. In combination, these choices lower latency and energy while maintaining modeling accuracy. 


\noindent\textbf{Main Contribution.} 
In this work we: (1) present a concurrent GRU mapping on FPGA that co-schedules computations across LUT and DSP resources with optimized data layout to accelerate the GRU-based MR; (2) improve on-chip memory efficiency through BRAM banking to sustain higher throughput under strict memory budgets; and (3) demonstrate MERINDA in an edge-AI setting using a real Automated Insulin Delivery (AID) application, evaluating accuracy, power, timing, and DRAM footprint while comparing against SINDY~\cite{zhang2019convergence}, PINN+SR~\cite{LiuPGLoss}, and LTC baselines.

\noindent\textbf{Novelty.} 
We co-design memory and compute: banked on-chip BRAM supplies the exact per-cycle read/write ports; fused operators eliminate intermediate stores and loads; and computation is mapped to the most suitable resources---DSP slices for MAC-heavy paths and LUT/carry chains for light element-wise logic. Using accuracy-budgeted fixed-point widths, this design yields an Initiation Interval, \( \mathrm{II} \approx 1 \) streaming pipeline with deterministic latency and superior performance-per-watt at small batch sizes.

\noindent\textbf{Why FPGA for this GRU.}
GPUs are built for large, batched kernels; a GRU step, however, is many \emph{small} kernels per time step, which on GPUs incurs kernel–launch overhead, runtime scheduling jitter, and Streaming Multiprocessor(SM) under-utilization at low batch sizes. You also cannot tailor memory port counts or numeric width per op, and frequent trips to external DRAM dominate latency/energy. An FPGA, by contrast, wire the GRU into a spatial, deeply pipelined dataflow with no per-step launches: one setup, then continuous streaming.

\begin{figure}[h]
\centering
\includegraphics[width=0.6\columnwidth]{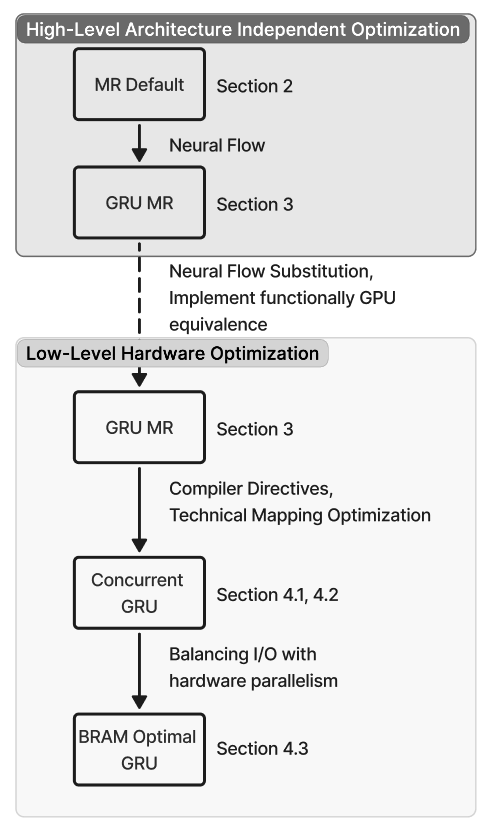} 
\caption{Optimization Framework of MERINDA.} 
\label{fig:Optimization_framework}
\end{figure}




\section{Overview of Optimization Framework}
Figure~\ref{fig:Optimization_framework} summarizes MERINDA’s two-layer optimization framework. 
At the high level, we reformulate Model Recovery (MR) by replacing the iterative ODE solver in LTC with a GRU-based neural flow. This preserves modeling accuracy while converting MR into a feed-forward, streaming computation that is well-suited for FPGA execution. 

At the low level, we specialize the GRU-based MR for FPGA by mapping linear operations to DSP slices and nonlinear element-wise functions to LUT/carry chains, enabling fine-grained spatial parallelism. In the Concurrent GRU design, we pipeline GRU stages across distinct on-chip memory resources and further boost throughput by banking BRAM to provide more read/write ports, allowing multiple pipeline stages to be fed in parallel. Together, these optimizations transform MR into a deeply pipelined hardware dataflow that preserves accuracy, minimizes latency, and sustains high throughput for training and inference.

\section{Background and Related Work}
We review the theoretical foundations of Model Recovery (MR) and summarize the current progress in MR acceleration, along with its practical applications.

\subsection{Theoretical background}
This section introduces the fundamentals of MR and establishes the approximate equivalence between neural flow architectures and NODEs.

\paragraph{Basics of Model Recovery}
The primary objective of MR is similar to an auto-encoder, where given a multivariate time series signal $X(t)$, the aim is to find a latent space representation that can be used to reconstruct an estimation $\Tilde{X}(t)$ with low error. It has the traditional encoder $\phi(t)$ and decoder ($\Psi(t)$) of an auto-encoder architecture\cite{baldi2012autoencoders}. MR represents the measurements $X$ of dimension $n$ and $N$ samples as a set of nonlinear ordinary differential equation model in (\ref{eqn:Model}). 
\begin{equation}
    \label{eqn:Model}
    \scriptsize
    \dot{X} = h(X,U,\theta),
\end{equation}
\vspace{-0.2in}\\
where $h$ is a parameterized nonlinear function that defines the hidden state dynamics, $U$ is the $m$ dimensional external input, and $\theta$ is the $p$ dimensional coefficient set of the nonlinear ODE model.

\noindent{\bf Sparsity:} An $n$-dimensional model with $M^{th}$ order nonlinearity can utilize $\binom{M+n}{n}$ nonlinear terms. A sparse model only includes a few nonlinear terms $p << \binom{M+n}{n}$. Sparsity structure of a model is the set of nonlinear terms used by it~\cite{hastie2015statistical}.

\noindent{\bf Identifiable model:} A model in (\ref{eqn:Model}) is identifiable~\cite{verdiere2019systematic}, if $\exists$ time $t_I > 0$, such that $\forall \theta, \Tilde{\theta} \in \mathcal{R}^p$:
\begin{equation}
    \label{eqn:Ident}
    \scriptsize
    \forall t \in [0,t_I], f(X(t),U(t),\theta) = f(X(t),U(t),\Tilde{\theta}) \implies \theta = \Tilde{\theta}. 
\end{equation}

\noindent (\ref{eqn:Ident}) effectively means that a model is identifiable if two different model coefficients do not result in identical measurements $X$. In simpler terms, this means $\forall \theta_i \in \theta, \frac{dX}{d\theta_i} \neq 0$. We assume that the underlying model is identifiable, where $f$ denotes a general nonlinear function.

\begin{problem}[Sparse Model Recovery]\label{prob:Problem} Given $N$ samples of measurements $X$ and inputs $U$, obtained from a sparse model in (\ref{eqn:Model}) such that $\theta$ is identifiable, recover $\Tilde{\theta}$ such that for $\Tilde{X}$ generated from $f(X,U,\Tilde{\theta})$, we have $||X - \Tilde{X}|| \leq \epsilon$, where $\epsilon$ is the maximum tolerable error.
\end{problem}

\begin{figure*}[h]
    \centering
        \includegraphics[width=\textwidth]{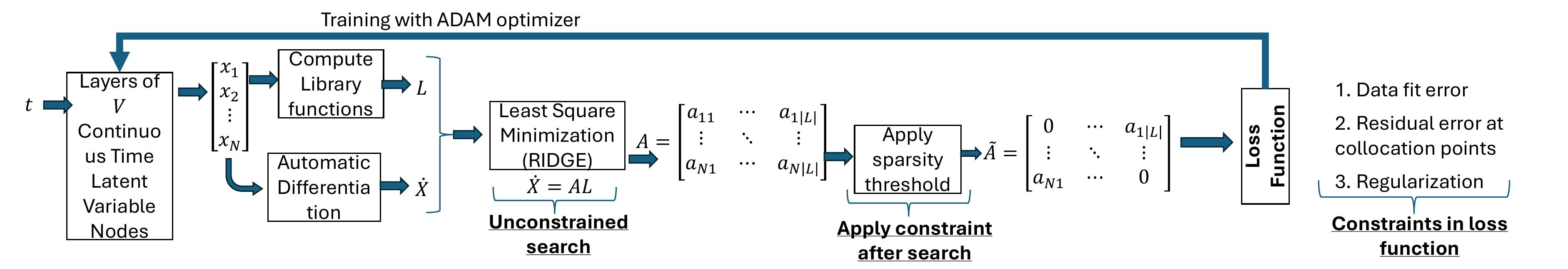}
    \caption{Generic architecture for physics-guided inferencing to be used in physical AI.}
    \label{fig:Arch}
\end{figure*}

\noindent{\bf State-of-the-Art:} Existing MR methods such as PINN+SR and PiNODE (Fig.~\ref{fig:Arch}) express $f$ as a linear combination $\dot{X}=AL$ of nonlinear functions from a library $L$, where $A$ is a coefficient matrix. To build the library, a fully connected continuous-time latent variable (CTLV) layer (e.g., Neural ODE~\cite{PiNode}) converts time $t$ into state estimates $X$. An automatic differentiation (autodiff) layer then estimates derivatives of $X$. Ridge regression identifies matrix $A$~\cite{kaiser2018sparse}, optimizing a loss function that balances data-fit error (real vs. reconstructed data via ODE solver), sparsity (regularization), and residual errors. The main computational costs arise from: i) the CTLV layer's ODE solver, ii) the autodiff RELU layer, and iii) Ridge regression.

\noindent{\bf Role of NODE:} Both EMILY~\cite{pmlr-v255-banerjee24a} and PINN+SR~\cite{chen2021physics} utilize a layer of NODE cells in order to integrate the underlying nonlinear ODE dynamics. NODE is one type of CTLV cell. NODE cell's forward pass is by design the integration of the function $h$ over time horizon $T$ with $N$ samples. This effectively requires an ODE solver in each cell of the NODE layer:
\vspace{-0.1in}\\
\begin{equation}
\scriptsize
z(t) = \int\limits^T_0{h(z,u,\theta)dt}, 
\end{equation}
where $z \in Z$ and $u \in U$ are each cells output and input, respectively. 

\paragraph{Neural flows and equivalent architectures to NODE}

According to the theory of neural flows~\cite{bilovs2021neural}, the node layer can be replaced by an approximate solution to $F(t) \approx Z(t)$ in discretized form using recurrent nerual network architectures such as GRU provided that following conditions are satisfied:

\begin{scriptsize}
\begin{equation}
    F(0,u) = Z(0,u), \text{(initial condition), and  }
    F(t,u) \text{  is invertible}.
\end{equation}
\end{scriptsize}

Bilovs et al.~\cite{bilovs2021neural} demonstrated that \( F(t,u) \) can be approximated by replacing the original NODE layer with a GRU layer. However, the GRU layer does not satisfy the invertibility condition, which is essential for ensuring the reconstruction of latent dynamics. To address this, the authors in~\cite{bilovs2021neural} propose the addition of a dense layer, leveraging its property as a universal approximator of nonlinear functions, thereby enabling it to approximate the inverse of \( F(t,u) \).  

MERINDA builds upon and enhances the equivalent architecture proposed in~\cite{bilovs2021neural} by further pruning the dense layer, as illustrated in Fig. \ref{fig:Approach}. The key idea is to optimize the dense layer structure by exploiting the inherent sparsity in the underlying model of the data, thereby improving computational efficiency without compromising accuracy.

\subsection{Related Work}
\subsubsection{Model Recovery}
The main aim of MR is to enhance Human-Centered Autonomous Systems (HCAS) risk management by allowing sufficient time for the \textit{human-in-the-loop (HIL)} to execute mitigative actions. Following the HIL control design model introduced by Seshia et al~\cite{SynthesisSeshia} and also adopted in our prior work~\cite{Banerjee2024}, we assume an error with inception at time $t=0$ takes time $t_h$ to evolve into a safety/security hazard (violation of property specified using Signal Temporal Logic(STL),  $\phi(Y)$). Assuming that the human reaction time is $t_r$ and the time taken to mitigate the violation of $\phi(Y)$ using mitigative action $a$ is $t_a$, the MR time overhead $t_{U2}$ should satisfy the constraint $t_{U2} \leq t_h - t_r - t_a$. This constraint is application dependent and can range from $\mu s$ in aviation, $ms$ in Autonamous Vehicle(AV) to $mins$ in Automated Insulin Delivery(AID). 
Deployments of HCAS are often resource constrained and energy / time costs of high dimensional ML/MR tehcniques may be prohibitive. While graphics processing units (GPUs) or tensor processing units (TPUs) are available in some HCAS such as AVs, it is not common. We want MR methods to have low processing overhead since: a) competition for available computation resource may interfere with essential perception functions, and b) MR may be stalled resulting in poor responsiveness. Hence, a dedicated low power and fast Application-Specific Integrated Circuit (ASIC) may be a viable if not optimal option for real time MR.
\begin{table}[t]
\centering
\scriptsize
\begin{tabular}{|p{0.35in}p{0.975in}p{0.03in}p{0.35in}|}
\hline
\textbf{Works} & \textbf{Platform / Networks} & \textbf{PG} & \textbf{Train/Test} \\
\hline
\multicolumn{4}{|c|}{New computation architecture} \\
\hline
\cite{hong2023physics} & GPU & No & Both \\
\hline
\cite{li2019rnn} & FPGA & No & Both \\
\hline
\cite{que2021accelerating,que2020reconfigurable,gao2020edgedrnn} & FPGA & No & Test \\
\hline
\multicolumn{4}{|c|}{Hybrid computation architecture} \\
\hline
\cite{cho2021farnn,ajili2022multimodal,liu2018hybrid} & GPU, FPGA & No & Both \\
\hline
\cite{he2021enabling} & GPU, FPGA & No & Train \\
\hline
\cite{jiang2021fleetrec} & GPU, FPGA & No & Test \\
\hline
\multicolumn{4}{|c|}{Novel network structures} \\
\hline
\cite{petersen2024convolutional} & Logic Gate NN & No & Both \\
\hline
\cite{hoefler2021sparsity,liu2022the,liu2023don} & Sparse NN & No & Both \\
\hline
\cite{qin2020binary,gholami2022survey,choi2018pact,rastegari2016xnor} & Binary/Quantized NN & No & Both \\
\hline
\multicolumn{4}{|c|}{Other hardware approaches} \\
\hline
\cite{zhao2019long,qu2022resonance} & Analog & No & Both \\
\hline
\end{tabular}
\caption{Related works on hardware acceleration of neural networks (NN) and physics-guided (PG) NN. Some techniques accelerate only test, while others accelerate both training and testing.}
\label{tab:related_work}
\end{table}

\begin{table}[t]
\centering
\scriptsize
\begin{tabular}{|l|c|c|c|}
\hline
\textbf{System} & \textbf{Time (s)} & \textbf{Energy (J)} & \textbf{DRAM (MB)} \\ \hline
AID System      & 56.63  & 107.88   & 192.36 \\ \hline
Autonomous Car  & 21.23  & 40.44    & 213.00 \\ \hline
APC System      & 20.74  & 39.43    & 289.18 \\ \hline
\end{tabular}
\caption{FPGA execution time, energy consumption, and DRAM footprint for different systems.}
\label{tab:comparison}
\end{table}

\noindent{\bf Gaps in SOTA:} Integration of Model Learning(ML) and Model Recovery(MR) necessitates simultaneous acceleration of both deep NNs such as transformers and physics-driven NNs. There are three types of approach towards accelerating deep NNs (Table \ref{tab:related_work}): a) \textit{using new computation architecture}, which implement the entire NN architecture in hardware accelerators such as FPGA, or analog circuits, b) \textit{using hybrid computation architectures}, where parts of NN is implemented in GPU, while computationally intensive components are accelerated in hardware, and c) \textit{defining novel network structure} that can be more efficiently accelerated such as logic gate networks, sparse NNs, quantized NNs, binary NNs. Table \ref{tab:related_work} shows that very little focus has been given to accelerating physics-driven NNs such as neural ODEs (NODE) or liquid time constant networks (LTC-NN) or SINDY-MPC, which form the backbone of the proposed MR solution. Acceleration of iterative operations in ODE solutions are critical hurdles. The forward pass of PINNs and reconstruction loss may need ODE solutions resulting in significant slowdown or heavy resource usage.

Table \ref{tab:comparison} shows the MR performance on a PYNQ-Z2 board with a 1.3M-configurable-gate FPGA using SINDY for the three running example systems. While MR performance is satisfactory for the AID ($t_{U2}$ > 5 mins), for the AV and APC systems quantized version of SINDY is not adequate. Moreover, we have also accelerated PINN with sparse regression~\cite{chen2021physics} in FPGA using quantization method (Table \ref{tab:related_work}) and achieve a speedup of 1.9 times with time requirement of 4 mins 10 seconds to learn the AID system.

\subsubsection{Optimization on FPGAs}
Most prior work tunes how a fixed kernel is placed and routed; we first change what the kernel is and how it streams. Instead of relying on mapping tools (e.g., Lakeroad) to discover better DSP/LUT layouts for an unchanged design~\cite{smith2024lakeroad}, we reshape the kernel and its memory schedule so it is mapping-friendly from the start. Rather than pushing frequency mainly via partitioning and floorplanning (e.g., RapidStream/RapidStream-IR)~\cite{guo2022rapidstream,lau2024rapidstreamir}, we pre-build a staged GRU pipeline and bank on-chip buffers to deliver exactly the reads/writes needed each cycle. 
While placement tools like OpenPARF model the FPGA’s heterogeneous fabric to ease routing~\cite{mai2023openparf}, we bank the data across multiple BRAMs and align them with the FPGA’s physical tile and column grid, then cluster the compute units—DSP slices for linear function and LUT/carry logic for non-linear functions—right next to these BRAM banks. This shortens wire lengths and makes place-and-route (P\&R) simpler and more predictable.
In sum, algorithm-memory-microarchitecture co-design removes iterative dependencies, matches per-cycle memory bandwidth to compute lanes, and cleanly partitions arithmetic across DSPs and LUT.

\section{MERINDA Architecture}

\begin{figure}[h]
\centering
\includegraphics[width=\columnwidth]{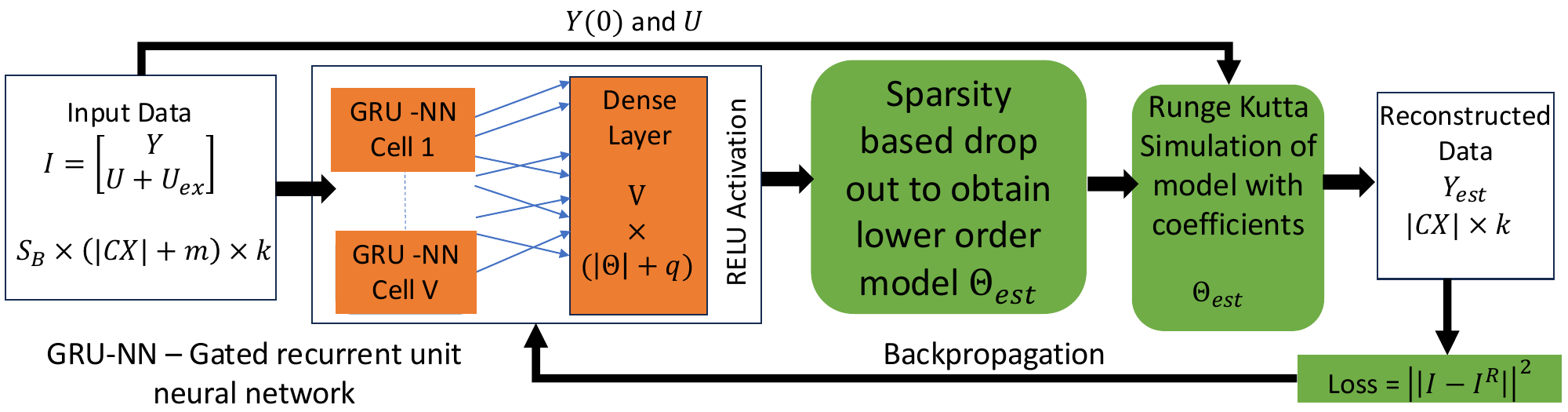} 
\caption{MERINDA: GRU NN-based MR architecture.} 
\label{fig:Approach}
\end{figure}

In our approach (Fig.~\ref{fig:Approach}), we extend the GRU architecture to design MERINDA, a neural model tailored for MR. The GRU’s forward pass encodes model coefficients as nonlinear functions of the inputs \( U \) and outputs \( Y \), converting the system dynamics into an overdetermined set of nonlinear equations. To handle redundancy and ensure consistent coefficient estimation, we introduce a dense layer to select a representative subset of equations. This selection is optimized using an \textit{ODE loss}—the mean squared error (MSE) between the observed \( Y \) and the predicted output \( Y_{\text{est}} \), computed by solving the system with an ODE solver: \( \mathbf{SOLVE}(Y(0), \Theta, U) \).

The advanced neural architectures for model recovery in Fig.~\ref{fig:Approach} are implemented by extending the base code available in~\cite{liquid-time-constant-networks}. The training data consists of temporal traces of \( Y \) and \( U \), where \( Y \) is sampled at a rate at least equal to the Nyquist rate for the given application, and \( U \) is sampled at the same rate as \( Y \). The resulting dataset is then divided into batches of size \( S_B \), forming a three-dimensional tensor of size \( S_B \times (|Y|+m) \times k \).

Each batch is processed through the \textit{GRU-NN} network with \( V \) nodes, generating \( V \) hidden states. A dense layer is then applied to transform these \( V \) hidden states into \( p = |\Theta| \) model coefficient estimates and \( q \) input shift values. The dense layer is structured as a multi-layer perceptron with a ReLU activation function for the model coefficient estimate nodes, whose outputs correspond to the estimated model coefficients. The dense layer maps the \( V \)-dimensional hidden state outputs to \( \binom{M+|X|}{|X|} \), which represents the number of nonlinear terms available for an \( M \)th-order polynomial. A dropout rate of \( |\Theta| \) is applied to ensure that the final number of active output layers corresponds to \( |\Theta| \).  

The estimated model coefficients, along with the initial value \( Y(0) \), are passed through an ODE solver to compute the solution for the nonlinear dynamical equations using the estimated coefficients \( \Theta_{\text{est}} \), initial conditions \( Y(0) \), and inputs \( U \). The solver employs the Runge-Kutta integration method to obtain \( Y_{\text{est}} \). During the backpropagation phase, the network loss is augmented with the ODE loss, defined as the MSE between the original trace \( Y \) and the estimated trace \( Y_{\text{est}} \).

\subsection{Proof of Equivalence}
We now prove that the architecture in Fig. \ref{fig:Approach} is equivalent to the NODE based MR architecture in Fig. \ref{fig:Arch}.

\noindent\textbf{Neural ODE Formulation.}  
In a NODE network, differentiating the forward pass of a single cell yields the standard ODE form
\begin{scriptsize}
\begin{equation}
\frac{dx(t)}{dt} = h( x(t),u,\theta).
\label{eq:node_diff}
\end{equation}
\end{scriptsize}

\medskip
\noindent\textbf{Koopman Operator Formulation.}  
According to Koopman theory~\cite{sinha2020koopmanoperatormethodsglobal}, the nonlinear dynamics in \eqref{eq:node_diff} can be represented in a potentially infinite-dimensional linear space. This representation involves:  
(i) a Koopman operator $K$,  
(ii) a measurement function $G$, and  
(iii) a set of latent variables $Z$.  
The dynamics are expressed as
\begin{scriptsize}
\begin{equation}
K\,G(t, Z) = \frac{d}{dt} G(t, Z), 
\qquad 
x(t) = G(t, Z).
\label{eq:koopman}
\end{equation}
\end{scriptsize}

\medskip
\noindent\textbf{Mori--Zwanzig (MZ) Formulation.}  
The Koopman representation can be partitioned into two components:  
(a) observable dynamics $G_m$, and  
(b) implicit dynamics $G_i$.  
The coupled system is then written as
\begin{scriptsize}
\begin{equation}
\frac{d}{dt}
\begin{bmatrix}
G_m \\
G_i
\end{bmatrix}
=
\begin{bmatrix}
K_M & K_{IM} \\
K_{MI} & K_I
\end{bmatrix}
\begin{bmatrix}
G_m \\
G_i
\end{bmatrix},
\label{eq:mz_coupled}
\end{equation}
\end{scriptsize}where $K_M$ and $K_I$ represent the self-dynamics of the observable and implicit states, while $K_{IM}$ and $K_{MI}$ capture cross-coupling effects.  

By applying the MZ projection formalism~\cite{hijon2010mori} and Laplace transform, the observable dynamics $G_m$ evolve as
\begin{scriptsize}
\begin{equation}
\frac{dG_m}{dt} 
= K_M G_m 
+ K_{IM} \int_{-\infty}^{t} e^{(t-s)K_I}\, K_{MI}\, G_m(s)\, ds ,
\label{eq:mz_gm}
\end{equation}
\end{scriptsize}where the first term captures the instantaneous effect of observables, while the second term introduces a \emph{memory effect}, expressed as a convolution between the kernel $e^{(t-s)K_I}K_{MI}$ and the past trajectory $G_m(s)$.  
For clarity, stochastic noise terms in the full MZ decomposition are omitted. Once $G_m$ is obtained, the original system state can be approximately reconstructed via a projection operator $P$:
\begin{scriptsize}
\begin{equation}
x(t) \approx P\,G_m(t),
\label{eq:mz_state}
\end{equation}
\end{scriptsize}where $P$ denotes the mapping from observable coordinates back to the physical state space.

\medskip
\noindent\textbf{GRU Formulation.}  
In a GRU cell, the hidden state is defined as
\begin{scriptsize}
\begin{equation}
h(t) = r(t) \circ h(t-1) + \big( 1 - r(t) \big) \circ c(t-1),
\label{eq:gru_update}
\end{equation}
\end{scriptsize}where $r(t)$ is the reset gate, $c(t)$ is the candidate state, and $\circ$ denotes elementwise multiplication.  
Rearranging, we obtain
\begin{scriptsize}
\begin{equation}
h(t) = h(t-1) + \big( 1 - r(t) \big) \circ \big(c(t-1) - h(t-1)\big).
\label{eq:gru_rearranged}
\end{equation}
\end{scriptsize}

The recurrence in \eqref{eq:gru_rearranged} has the same structure as the Mori--Zwanzig approximation in \eqref{eq:mz_gm}, where the convolutional memory term models the implicit dynamics. Hence, for every NODE-based architecture, there exists an equivalent GRU-based formulation with parameters that capture the effects of both observable and implicit dynamics.

\section{Hardware Design}
In this section, we present the hardware architecture of MERINDA, detailing the design principles, dataflow organization, and key optimizations that enable efficient FPGA acceleration of Model Recovery.

\subsection{Overall Architecture}

\begin{figure}[t]
\centering
\includegraphics[width=0.95\linewidth]{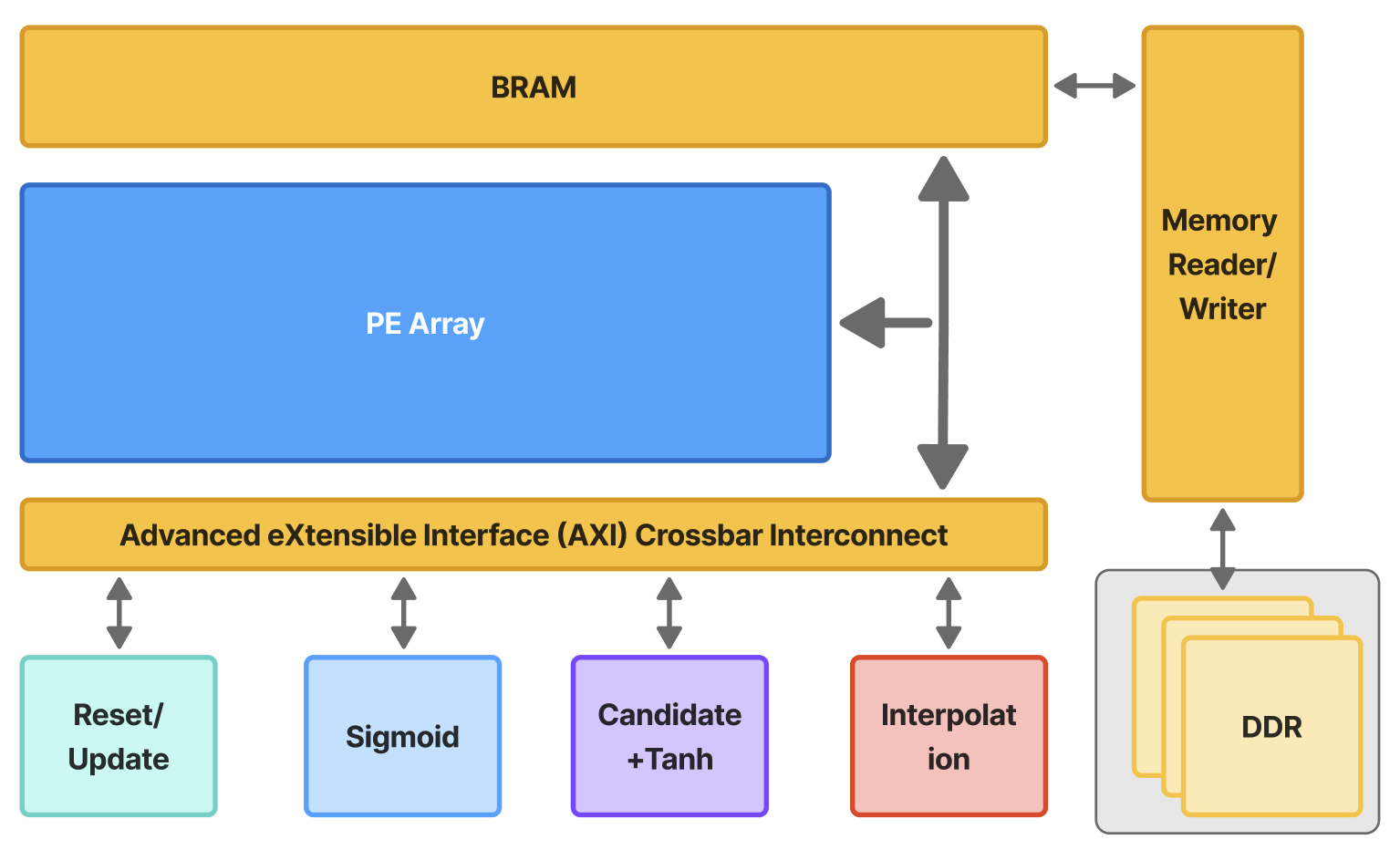}
\caption{
Overall GRU accelerator architecture. The \textbf{PE Array} is backed by on-chip \textbf{BRAM} and connected to a \textbf{Memory Reader/Writer} that streams data to/from off-memory (\textbf{DDR}). 
}
\label{fig:arch_opt}
\end{figure}


Unlike LTC that require iterative numerical solvers, the GRU performs a fixed sequence of linear and nonlinear operations that can be directly mapped onto FPGA hardware without additional ODE-solving logic.
Since the forward pass dominates the computation in both GRU training and inference—repeated at every time step for long sequences—its efficient implementation is the key target for hardware acceleration.
Modern FPGAs offer two complementary primitives: (i) \textbf{LUT logic}, which can be used as combinational logic or small ROMs for table lookups, and (ii) \textbf{DSP slices} (e.g., DSP48E2), which provide pipelined fused multiply–accumulate (MAC) datapaths.
In our GRU accelerator, we assign \emph{nonlinear activations} to LUT/ROM tables and \emph{all linear arithmetic} (dot products, elementwise scaling, and blending) to DSP slices.

Fig.~\ref{fig:arch_opt} summarizes the overall architecture.
A central PE Array executes the core compute kernels while on-chip (BRAM) buffers activations/weights to minimize off-chip traffic.
A dedicated Memory Reader/Writer engine streams inputs and parameters from off-chip (DDR) into BRAM and writes back results through AXI master ports.
Specialized functional units—Reset/Update, Sigmoid, Candidate+Tanh, and Interpolation—implement the nonlinearities and linearities on DSP and LUT.
All modules communicate over an AXI Crossbar interconnect, which arbitrates shared-resource access and enables concurrent dataflow between the PE Array, functional units, BRAM, and the memory subsystem.

\noindent\textbf{LUTs for nonlinearities\cite{hajduk2017high}.}
Sigmoid and $\tanh$ are fixed, elementwise nonlinear functions. Implementing them arithmetically would require iterative exponentials/divisions and many multipliers. Instead, we use LUT to produce activations in constant time (one cycle) with minimal DSP usage. Note that $(1 - z_t)$ in Eq.\ref{eq:gru_update} itself is linear and does not require a nonlinear LUT; only $z=\sigma(\cdot)$ and $\tanh(\cdot)$ do.

\noindent\textbf{DSPs for linear arithmetic.}
All weighted sums and blends are MAC-dominant and map directly to DSP slices at II$=1$. For example,

\begin{align}
r_t &= \sigma\!\bigl(W_r x_t + U_r h_{t-1} + b_r\bigr), \label{eq:gru_r}\\
z_t &= \sigma\!\bigl(W_z x_t + U_z h_{t-1} + b_z\bigr), \label{eq:gru_z}\\
\tilde{h}_t &= \tanh\!\bigl(W_h x_t + U_h\bigl(r_t \odot h_{t-1}\bigr) + b_h\bigr). \label{eq:gru_h}
\end{align}

where each matrix–vector product $W_{\{\cdot\}}x_t$ and $U_{\{\cdot\}}(\cdot)$ is realized by arrays of DSP MAC lanes; bias adds are absorbed in the DSP post-adders, and the resulting affine outputs are then passed to lightweight nonlinear units such as $\sigma(\cdot)$ or $\tanh(\cdot)$ for gate activation, where $r_t$ decides \emph{what past information to forget}, $z_t$ decides \emph{how much to update}, and $\tilde h_t$ is the \emph{new information proposed to enter the hidden state}.

The final interpolation
\begin{equation}
h_t = (1 - z_t)\odot \tilde{h}_t + z_t \odot h_{t-1}.
\label{eq:gru_update}
\end{equation}
is two elementwise multiplies plus one add, again a natural fit for DSPs. The performance-critical accumulations stay in DSPs.

\subsection{GRU Acceleration Flow}

The GRU forward pass combines two fundamentally different types of computation:
\emph{linear arithmetic} (matrix--vector multiplications, bias additions, interpolations)
and \emph{nonlinear activation functions} (sigmoid, tanh).
FPGAs provide two complementary compute primitives—\textbf{DSP slices} and \textbf{LUT logic}—that align naturally with these operations
and can be orchestrated concurrently under \texttt{DATAFLOW}, which is an HLS compiler directive.

\begin{figure}[htbp]
\centering
\includegraphics[width=\columnwidth]{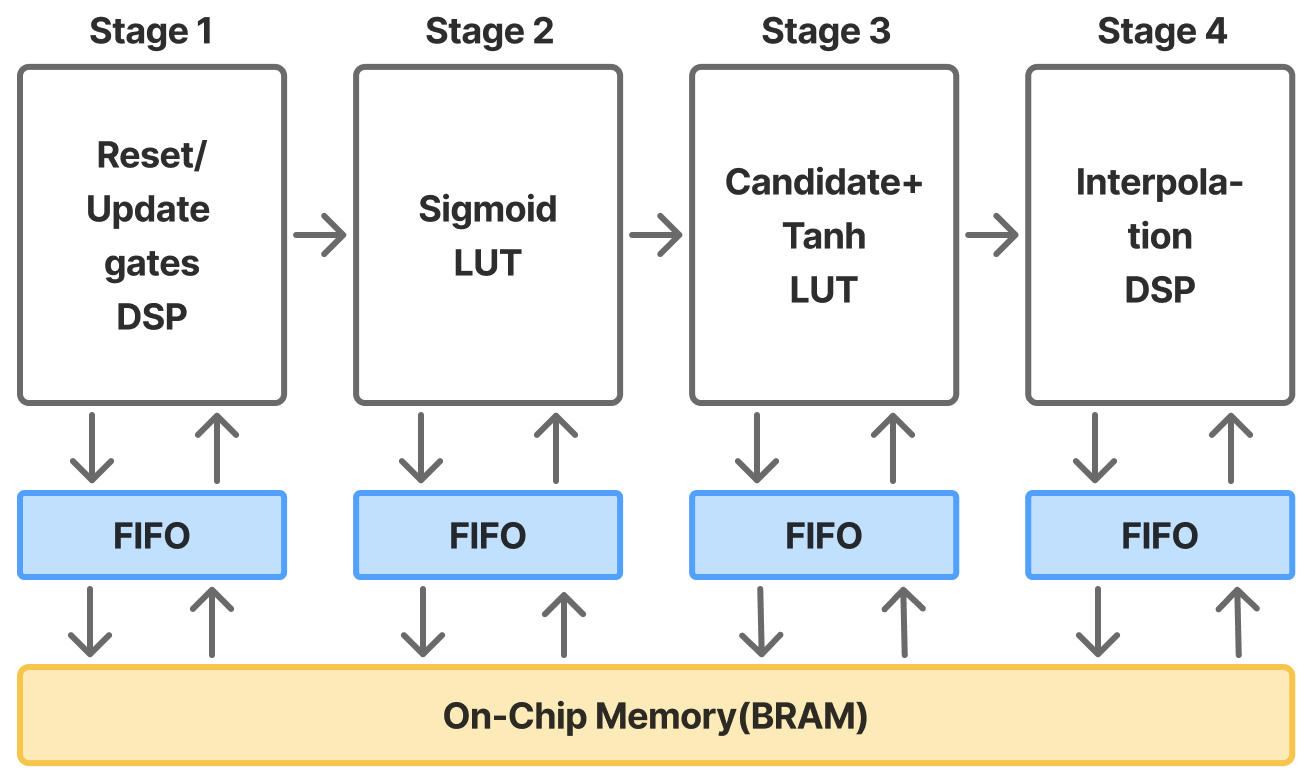}
\caption{GRU forward-pass pipeline on FPGA. Stage~1: gate affines (DSP); Stage~2: sigmoid (LUT); Stage~3: candidate (DSP + tanh LUT); Stage~4: interpolation (DSP). Stream FIFOs decouple stages, and a global On-Chip memory (BRAM) sustains II=1 by overlapping load/compute/store.}
\label{fig:data_mapping}
\end{figure}

\subsubsection{DSP slices accelerate linear GRU operations}  
DSP blocks on modern FPGAs (e.g., Xilinx DSP48E2) contain (1) a fixed-latency hardware multiplier, (2) a fused multiply–add datapath, for example $P = A \times B + C$, and (3) dedicated pipeline registers that enable sustained II = 1 operation at several hundred MHz.
GRU stages involve repeated dense linear operations:
\[
W_r x_t,\quad U_r h_{t-1},\quad
W_z x_t,\quad U_z h_{t-1},\quad
W_h x_t,\quad U_h (r_t\odot h_{t-1})
\]
followed by bias addition and blending.
Here, $x_t$ is the input vector at time step $t$, and $h_{t-1}$ is the hidden state from the previous time step. 
The matrices $W_r$, $W_z$, and $W_h$ are the input–to–gate weight matrices for the reset, update, and candidate gates, respectively. 
The matrices $U_r$, $U_z$, and $U_h$ are the recurrent (hidden–to–gate) weight matrices that capture the influence of the previous hidden state. 
The term $r_t\odot h_{t-1}$ denotes the elementwise modulation of the previous hidden state by the reset gate, ensuring that only selected past information contributes to the candidate computation. 

After these linear transformations, gate-specific bias vectors are added to each pre-activation, and the results are passed through nonlinear activations (sigmoid for $r_t$ and $z_t$, and $\tanh$ for $\tilde h_t$) before the final interpolation which shown in Eq.\ref{eq:gru_update}.

Each Multiply–Accumulate (MAC) lane in these matrix–vector products maps directly onto a DSP slice so that every clock one product is computed and accumulated.
Because DSPs are hard-wired, they avoid the overhead of implementing multipliers or wide adders in LUT fabric, yielding both higher clock frequency and lower power.

\subsubsection{LUT logic accelerates nonlinear GRU operations.}  

The gates require nonlinear activations:
\[
r_t = \sigma(\cdot),\qquad
z_t = \sigma(\cdot),\qquad
\tilde h_t = \tanh(\cdot),
\]
whose mathematical definitions involve expensive functions (exponentials and divisions).
On FPGA, these are implemented as \emph{look-up tables or piecewise-linear tables} stored in distributed LUT RAM or small BRAMs.
Such tables return the nonlinear output in \emph{constant time} (one cycle) instead of requiring iterative evaluation.
Since the nonlinear functions are fixed and low-dimensional (element-wise), LUTs can serve hundreds of parallel gate units without consuming DSPs.

\subsubsection{Synergy of DSP and LUT for concurrent stages.}  
The separation between linear and nonlinear computations allows them to execute in parallel on different resources:
DSP slices handle MAC-dominated matrix–vector products, while LUT blocks service nonlinear activations at the gate outputs.
In a dataflow architecture, Stage~1 (weighted sums in DSPs), Stage~2 (sigmoid in LUTs plus reset modulation in DSPs),
Stage~3 (tanh in LUTs plus candidate accumulation in DSPs),
and Stage~4 (update-gate blending in DSPs) can each become a distinct \emph{pipeline stage} connected by BRAM FIFOs.

The \texttt{DATAFLOW} pragma tells the HLS compiler to generate separate hardware processes for each stage. Once the pipeline is filled, every stage operates on a different time step $t$ of the GRU sequence in the same clock cycle: \textbf{Stage~1} computes gate accumulations for time step $t{+}1$ using DSPs, \textbf{Stage~2} applies sigmoids and reset modulation for time step $t$ using LUTs and DSPs, \textbf{Stage~3} generates the candidate hidden state for time step $t{-}1$, and \textbf{Stage~4} blends the final hidden state for time step $t{-}2$.
Thus DSP-intensive and LUT-intensive computations proceed concurrently without competing for the same resources.

\subsection{Optimization of throughput}

\begin{figure}[htbp]
\centering
\includegraphics[width=\columnwidth]{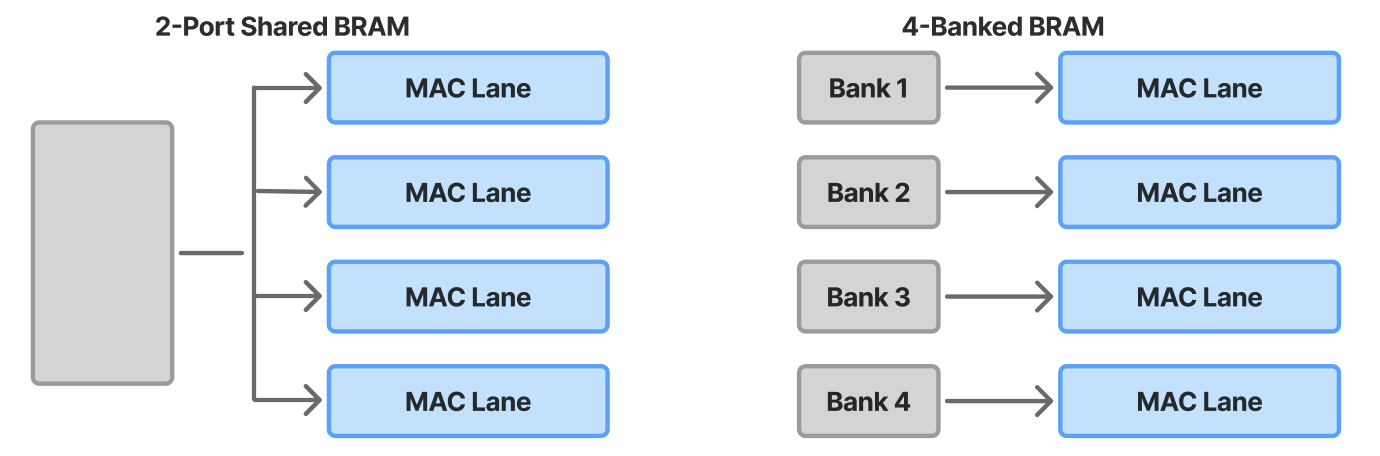}
\caption{Left (a): Single dual-port BRAM provides only two ports, so 4 Multiply–Accumulate Calculation(MAC) lanes require two cycles ($\text{II}=2$). 
Right (b): 4-banked BRAM provides eight ports, allowing all 4 MAC lanes to be served in one cycle ($\text{II}=1$).}
\label{fig:bram_banking}
\end{figure}

Even with \texttt{DATAFLOW} enabled to overlap pipeline stages, the overall throughput is limited by the slowest stage.  
A common bottleneck is memory access: if a loop iteration needs several array reads in the same cycle (for example because of \texttt{UNROLL}), 
but the memory cannot supply them all, the pipeline stalls.

\subsubsection{Initiation Interval and BRAM Banking}
\noindent\textbf{Initiation Interval} measures how frequently a pipeline stage can launch new loop iterations.  
It is defined as the number of clock cycles between the start of two consecutive iterations in the same stage.  
A perfectly pipelined loop has $\mathrm{II}=1$, meaning a new iteration begins every clock cycle.  
If some resource (typically memory) cannot supply all required data in the same cycle, the stage must pause, and the II rises above~1.

Let $R$ be the number of memory reads required per loop iteration.  
A single true dual-port BRAM provides only two accesses per cycle, so if $R>2$, the loop cannot launch a new iteration every cycle:
\[
\mathrm{II} \;\ge\; \Big\lceil \frac{R}{2} \Big\rceil .
\]

\noindent\textbf{Banking} alleviates this bottleneck by splitting one large array into $B$ independent memory banks, each still dual-ported.  
This increases the number of parallel accesses to
\[
\text{ports per cycle} = 2B, \qquad
\mathrm{II} \;\ge\; \Big\lceil \frac{R}{2B} \Big\rceil .
\]
By choosing $B$ so that $2B \ge R$, the stage can fetch all required data each cycle and achieve the optimal $\mathrm{II}=1$, where the initiation interval (II) is the number of cycles between starting consecutive iterations, $\mathrm{II}=1$ means a new iteration launches every clock cycle, maximizing throughput.

For example, in a GRU gate with \texttt{UNROLL}=4, four parallel MAC lanes each fetch one weight per cycle, so $R=4$.  
With a single bank ($B=1$), only two ports are available and the stage must wait:
\[
\mathrm{II} \ge \lceil 4 / 2 \rceil = 2 \quad (\text{stalling}),
\]
whereas with two banks ($B=2$), four ports are available and
\[
\mathrm{II} \ge \lceil 4 / 4 \rceil = 1 \quad (\text{full throughput}).
\]
If the stage also needs to fetch four recurrent-matrix weights in the same cycle ($R=8$), at least $B=4$ is required to keep $\mathrm{II}=1$.

\noindent\textbf{Impact on DATAFLOW.}  
Banking not only improves the throughput of a single stage but also prevents contention between concurrent stages in \texttt{DATAFLOW}.  
Without sufficient banking, multiple stages compete for the same BRAM ports, forcing HLS to insert arbitration or time-multiplexing, which raises the effective II.  
Partitioning or replicating arrays gives each stage enough ports, and BRAM-based FIFOs between stages further decouple them and prevent stalls from propagating.

\subsubsection{Directives for BRAM Banking} 
The following HLS pragmas improve pipeline throughput in the GRU accelerator by aligning memory layout with parallel compute and by providing sufficient memory ports.

\noindent\textbf{Partitioning for parallel access.}  
Each GRU gate (\(r,z,h\)) performs a matrix–vector product.  
With \texttt{UNROLL}=4, four MAC lanes must fetch four weights each cycle.  
Partitioning splits the weight array into 4 independent BRAM banks (8 ports total), matching the unrolled lanes and avoiding port contention:
\begin{verbatim}
#pragma HLS ARRAY_PARTITION \
    variable=params.Wr dim=2 factor=4 cyclic
#pragma HLS BIND_STORAGE \
    variable=params.Wr type=rom_2p impl=bram
// repeat for Ur, Uz, Uh, Wz, Wh, Wy
\end{verbatim}

\noindent\textbf{Reshaping for wide-word access.}  
Packing four weights into a single wide word reduces addressing overhead and improves bandwidth efficiency:
\begin{verbatim}
#pragma HLS ARRAY_RESHAPE \
    variable=params.Wr dim=2 factor=4
\end{verbatim}

\noindent\textbf{BRAM-based FIFOs for DATAFLOW.}  
Different stages (e.g., gate computation, activation, state update) produce and consume vectors at different rates.  
BRAM FIFOs decouple them and prevent stalls from propagating:
\begin{verbatim}
#pragma HLS STREAM \
    variable=r_pre depth=256
#pragma HLS BIND_STORAGE \
    variable=r_pre type=fifo impl=bram
// repeat for z_pre, h_pre, r_gate, z_gate, h_candidate
\end{verbatim}

\texttt{ARRAY\_PARTITION} or \texttt{ARRAY\_RESHAPE} implements explicit BRAM banking: one large array is split into smaller sub-arrays, each with its own dual ports.  
This ensures all unrolled MAC lanes can read simultaneously and keeps the loop initiation interval at $\mathrm{II}=1$.  
\texttt{STREAM} with \texttt{BIND\_STORAGE(...impl=bram)} places FIFOs in dedicated BRAM blocks, effectively giving each \texttt{DATAFLOW} stage its own bank so they do not contend for ports.  

Thus, partitioning and reshaping create the banks that feed the parallel compute lanes, while streaming allocates separate BRAM banks for inter-stage buffers.  
Both are essential to avoid port bottlenecks and to sustain full pipeline throughput in the GRU accelerator.

\medskip
\noindent\textbf{Limitations of Excessive Banking.}  
Although increasing the bank factor $B$ provides more ports and can lower the II, very aggressive banking has drawbacks: (1) each bank consumes separate BRAM blocks, increasing overall BRAM usage and possibly exceeding device capacity; (2) extra banks increase routing complexity and can raise critical-path delay, potentially lowering the maximum clock frequency; and (3) large numbers of banks add address-decoding and fan-out overhead, which may offset the expected throughput gain. Hence, $B$ should be chosen to just meet the required parallel accesses ($2B \ge R$).

\section{Evaluation and Results}
We describe the implementation details and platform setup, followed by an analysis of MERINDA’s performance improvements.

\subsection{Case Studies and Data}\label{AA}
\noindent{\bf Simulation case studies:} Chaotic Lorenz and F8 Cruiser case studies were obtained from~\cite{kaiser2018sparse} and data was generated by implementing the models in Matlab and using ODE45 to solve the ODEs. 

\noindent{\bf Real world case studies:} Following datasets were used.


\noindent {\bf Lotka Volterra:} We used yearly lynx and hare pelts data collected from Hudson Bay Company~\cite{kaiser2018sparse}.

\noindent {\bf Pathogenic attack system:} The data is available in~\cite{kaiser2018sparse}.

\noindent {\bf Automated Insulin Delivery (AID):} The datasets are patient data obtained from the OhioT1D dataset available in~\cite{marling2020ohiot1dm}. It is 14 time series data of glucose insulin dynamics. Each time series had a duration of 16 hrs 40 mins which amounts to 200 samples of CGM and insulin.

\subsection{Experiment Setup}
To evaluate the performance of the FPGA, we conduct experiments on both the GPU and FPGA. The GPU serves as the baseline for performance comparison.  

\noindent{\bf GPU Platform:} The experiments were conducted on a system equipped with an Intel Xeon w9-3475X CPU and an NVIDIA RTX 6000 GPU with 48GB of memory. The model was implemented using TensorFlow 2.10 and Keras 2.10. Power consumption was monitored using {\tt nvidia-smi}, while execution time and DRAM footprint were measured using the {\tt time} and {\tt psutil} libraries.  

\noindent\textbf{Mobile GPU Platform:} Our experiments are conducted on the NVIDIA Jetson Orin Nano Developer Kit, which features a 6-core Arm Cortex-A78AE CPU and 8~GB of LPDDR5 memory. The integrated GPU is based on the NVIDIA Ampere architecture, equipped with 1024 CUDA cores and 32 Tensor Cores.

\noindent{\bf FPGA Platform:} For the FPGA implementation, the experiments were performed on a PYNQ-Z2 board featuring a Dual-Core ARM Cortex-A9 processor and a 1.3M-configurable-gate FPGA. The GRU model was developed from scratch, with the forward pass and backpropagation logic implemented in C++ using High-Level Synthesis (HLS) within AMD’s Vitis tool. The accelerator for the forward pass was integrated with the FPGA board using Direct Memory Access (DMA) to interface with the processing system on the PYNQ-Z2. DRAM footprint were recorded using the {\tt psutil} library.

\subsection{Power Measurement}
Power consumption was measured using the FPGA’s on-board sensors accessed via the \texttt{libsensors} API, with a Python-based DataRecorder sampling voltage, current, and power rails in real time. We align sampling with kernel start/stop, discard a short warm-up phase, and report averages over repeated runs.

We did not use an external power meter because external meters often sample too slowly for the short, bursty phases of MR training. They also lack per-rail visibility, making it hard to separate compute from memory power. On-board measurement is closer to the silicon, faster, and more specific to the accelerator, giving cleaner, reproducible measurements for our use case.

\begin{table*}[ht]
\centering
\scriptsize
\caption{Comparison across platforms for four workloads on AID dataset.}
\label{tab:img-derived-comparison}
\begin{tabular}{l
                ccc
                ccc
                ccc
                ccc
                ccc}
\toprule
& \multicolumn{3}{c}{\textbf{Average Error}} 
& \multicolumn{3}{c}{\textbf{Runtime (S)}} 
& \multicolumn{3}{c}{\textbf{Average Power (W)}} 
& \multicolumn{3}{c}{\textbf{DRAM Footprint}} 
& \multicolumn{3}{c}{\textbf{Freq (MHz)}} \\
\cmidrule(lr){2-4}\cmidrule(lr){5-7}\cmidrule(lr){8-10}\cmidrule(lr){11-13}\cmidrule(lr){14-16}
\textbf{Workload} 
& \textbf{FPGA} & \textbf{Mobile GPU} & \textbf{GPU}
& \textbf{FPGA} & \textbf{Mobile GPU} & \textbf{GPU}
& \textbf{FPGA} & \textbf{Mobile GPU} & \textbf{GPU}
& \textbf{FPGA} & \textbf{Mobile GPU} & \textbf{GPU}
& \textbf{FPGA} & \textbf{Mobile GPU} & \textbf{GPU} \\
\midrule
\textbf{LTC}
& 3.76  & 3.993 & 2.55
& 1755.04 & 1100.82 & 1152.92
& 5.10  & 6.855 & 81
& 494.62    & 2532.45 & 6465.07
& 177   & 306    & 1410 \\
\textbf{SINDY}
& 7.37  & 4.44  & 3.39
& 259.36 & 384.713 & 86.88
& 4.73  & 5.88  & 62
& 214.45    & 1040.8 & 4433.51
& 144   & 306    & 1410 \\
\textbf{PINN+SR}
& 8.02  & 4.43  & 3.396
& 256.90 & 389.36 & 85.05
& 4.81  & 5.45  & 64.01
& 229.14 & 1044.55 & 4377.54
& 165   & 306    & 1410 \\
\textbf{MR}
& 4.60  & 3.07  & 2.89
& 251.97 & 423.75 & 380.21
& 4.90 & 5.53 & 72
& 214.23    & 2355.13 & 6118.36
& 173   & 306    & 1410 \\
\bottomrule
\end{tabular}
\end{table*}

\subsection{Implementation Details}\label{AA}
We implemented the GRU accelerator in C/C++ using \textbf{Vitis~HLS} and integrated the generated RTL into Vivado for deployment via AXI-based DMA. The HLS kernel exposes \texttt{AXI4-Stream} interfaces to transfer input and output tensors between DDR memory and the accelerator through DMA, while on-chip BRAM is used to store model weights and to serve as stage-to-stage FIFOs within the accelerator.

Inside the kernel we use \texttt{DATAFLOW} to split the GRU forward pass into concurrent stages  
(gate accumulation, nonlinear activation, candidate computation, and final blending).  
Inner compute loops are pipelined with $\mathrm{II}=1$, while \texttt{ARRAY\_PARTITION}/\texttt{ARRAY\_RESHAPE} and  
\texttt{BIND\_STORAGE(impl=bram)} directives provision enough memory ports to keep the unrolled DSP MAC lanes fed.  
Stage-to-stage data is passed through \texttt{STREAM} FIFOs (mapped to BRAM) to decouple producer and consumer rates.  

Linear arithmetic (matrix–vector products, bias adds, blending) is mapped to DSP48 MAC units.  
Nonlinear gates such as sigmoid and $\tanh$ use LUT- or BRAM-based lookup or piecewise-linear tables, producing outputs in a single cycle without using DSPs.  
We use fixed-point (\texttt{ap\_fixed}) formats: 8–16\,b for activations and 12–16\,b for weights and accumulators to balance accuracy, resource use, and achievable clock frequency.

After C-simulation and C/RTL co-simulation, the synthesized IP is exported from Vitis HLS and integrated into a Vivado block design.  
The kernel connects via \texttt{S\_AXILITE} to the processor control bus and via \texttt{AXIS}/\texttt{M\_AXI} ports to DMA engines and external DDR.  
We typically drive the accelerator with a 150-200\,MHz PL clock, using a suitable stream width (e.g., 128\,bits) and burst-aligned DMA transfers.

On the host side (XRT runtime or bare-metal/Linux on Zynq), we program the FPGA bitstream, allocate pinned DDR buffers, enqueue DMA transfers for input $x_t$ sequences and the initial $h_0$, start the kernel, and collect accuracy, timing and DRAM footprint.  

\subsection{Results}
\subsubsection{Accuracy Benchmark} 
Compared to SOTA MR methods such as EMILY and PINN+SR, Table~\ref{tbl:SMRFM} shows that MERINDA achieves comparable or even lower errors across four benchmark applications. Accuracy is measured using Mean Square Error between the estimated parameters and the ground truth values. 

\noindent{\bf Key takeaway:} From Table \ref{tbl:SMRFM} we see that for all four benchmark applications, MERINDA architecture is successful in recovering the underlying dynamics with comparable accuracy as SOTA methods which use the standard Tensorflow pipeline for neural network training in GPU based systems.

\begin{table}[H]
\centering
\scriptsize
\caption{Comparison between MERINDA and SOTA MR techniques EMILY and PINN+SR using reconstruction MSE. Errors are absolute values; numbers in parentheses indicate standard deviation.}
\label{tbl:SMRFM}
\begin{tabular}{lccc}
\toprule
\textbf{Applications} & \textbf{EMILY} & \textbf{PINN+SR} & \textbf{MERINDA} \\
\midrule
Lotka Volterra      & 0.03 (0.02)     & 0.05 (0.03)     & 0.03 (0.018) \\
Chaotic Lorenz      & 1.7 (0.6)       & 2.11 (1.4)      & 1.68 (0.4) \\
F8 Cruiser          & 4.2 (2.1)       & 6.9 (4.4)       & 5.1 (2.2) \\
Pathogenic Attack   & 14.3 (12.1)     & 21.4 (5.4)      & 15.1 (10.2) \\
\bottomrule
\end{tabular}
\vspace{-0.1 in}
\end{table}

\begin{table}[h!]
\centering
\caption{Stage-wise compute mapping on FPGA. \(s_i\) denotes the \(i\)-th computation stage (see Fig.~\ref{fig:data_mapping}); \(D\) indicates DSP-based MACs, and \(L\) indicates LUT/carry-chain logic.}
\small
\begin{tabular}{lccccc}
\toprule
\textbf{Config} & \textbf{Cycles} & \textbf{LUT} & \textbf{FF} & \textbf{DSP} & \textbf{BRAM} \\
\midrule
s1D\_s2D\_s3D\_s4D & 390 & 17415 & 16190 & 185 & 10 \\
s1D\_s2D\_s3D\_s4L & 389 & 17355 & 16140 & 184 & 10 \\
s1D\_s2D\_s3L\_s4D & 388 & 20020 & 17025 & 163 & 10 \\
s1D\_s2D\_s3L\_s4L & 387 & 19963 & 16977 & 162 & 10 \\
s1D\_s2L\_s3D\_s4D & 393 & 16970 & 16400 & 191 & 10 \\
s1D\_s2L\_s3D\_s4L & 392 & 16908 & 16351 & 190 & 10 \\
\textbf{s1D\_s2L\_s3L\_s4D} & \textbf{380} & \textbf{19480} & \textbf{17150} & \textbf{168} & \textbf{10} \\
s1D\_s2L\_s3L\_s4L & 390 & 19563 & 17209 & 168 & 10 \\
s1L\_s2D\_s3D\_s4D & 392 & 31620 & 23075 & 61  & 10 \\
s1L\_s2D\_s3D\_s4L & 391 & 31563 & 23011 & 60  & 10 \\
s1L\_s2D\_s3L\_s4D & 390 & 34225 & 23895 & 39  & 10 \\
s1L\_s2D\_s3L\_s4L & 389 & 34171 & 23848 & 38  & 10 \\
s1L\_s2L\_s3D\_s4D & 388 & 31170 & 23280 & 67  & 10 \\
s1L\_s2L\_s3D\_s4L & 387 & 31116 & 23222 & 66  & 10 \\
s1L\_s2L\_s3L\_s4D & 386 & 33820 & 24110 & 45  & 10 \\
s1L\_s2L\_s3L\_s4L & 392 & 33771 & 24080 & 44  & 10 \\
\bottomrule
\end{tabular}
\label{tab:calibrated_results}
\end{table}

\subsubsection{Analysis on Hardware Efficiency}
Table~\ref{tab:img-derived-comparison} highlights the relative advantage of the \texttt{MR} workload across platforms on AID application. 
In terms of \textbf{accuracy}, \texttt{MR} achieves the smallest gap between FPGA and GPU (4.60 vs.\ 2.89), unlike \texttt{SINDY} or \texttt{PINN+SR}, where the FPGA lags much further behind. 
For \textbf{runtime}, \texttt{MR} is among the fastest on FPGA (352\,ms) compared to other workloads that exceed 250--1700\,ms, indicating that its compute pattern maps well to FPGA pipelines. 
In \textbf{power efficiency}, \texttt{MR} maintains the low FPGA power advantage (4.9\,W) while avoiding the long latencies seen in \texttt{LTC}. 
Its \textbf{memory footprint} (72\,MB on FPGA) is moderate and far smaller than the multi-GB usage on GPUs, supporting deployment in memory-limited edge devices. 

From Table~\ref{tab:calibrated_results}, we compare different combintations of stage-wise compute mappings for the GRU pipeline, where \(s_i\) indexes stages and each stage is realized either on DSP MACs (\(D\)) or LUT/carry-chain logic (\(L\)) under a fixed BRAM budget. Among all combinations, \(\textbf{s1D\_s2L\_s3L\_s4D}\) achieves the lowest cycle count (\(\mathbf{380}\)) while maintaining a balanced footprint across LUT/FF/DSP resources. This outcome matches our design intent: mapping arithmetic-intense stages to DSPs reduces critical-path depth, whereas using LUT/carry chains for lighter stages preserves DSP headroom and eases routing pressure.

\begin{table}[ht]
\centering
\scriptsize
\setlength{\tabcolsep}{4pt}
\renewcommand{\arraystretch}{0.85}
\caption{%
Cycle count, resource utilization, and interval for four FPGA designs: 
(1) \textbf{LTC (ODE) Baseline} – Liquid Time-Constant model using ODE solver, 
(2) \textbf{GRU Baseline} – conventional GRU forward pass without concurrency, 
(3) \textbf{GRU + \texttt{DATAFLOW} (Concurrent)} – GRU forward pass pipelined into stages with concurrent execution, and 
(4) \textbf{GRU + BRAM Banking (Optimized)} – further optimized GRU design with aggressive BRAM partitioning for maximum throughput.
}
\label{tab:gru-configs}
\begin{tabular}{lrrrrrrr}
\toprule
\textbf{Configuration} & \textbf{Cycles} & \textbf{LUT} & \textbf{FF} & \textbf{DSP} & \textbf{BRAM} & \textbf{Interval} & \textbf{Power (W)}\\
\midrule
LTC                   & 1201  & 27368   & 39281   & 49   & 5   & 12014 & 5.11 \\
GRU Baseline                          & 1045  & 10458   & 15538   & 44   & 7   & 271 &4.736  \\
Concurrent GRU  &   380  & 19480   & 17150   & 168  & 10  & 145 & 3.013  \\
BRAM optimal GRU        &   190  & 276047  & 130106  & 524  & 18  & 107  & 4.15 \\
\bottomrule
\end{tabular}
\end{table}


Table~\ref{tab:gru-configs} compares cycle counts and FPGA resources across four configurations. 
Moving from the LTC (ODE) baseline to the GRU baseline reduces the forward-pass time by $\approx 13.0\%$ (1201$\rightarrow$1045 cycles; speedup $=1201/1045\approx1.15\times$), reflecting the benefit of a fixed, feed-forward GRU over an iterative ODE solver. 
Introducing \texttt{DATAFLOW} concurrency (\emph{Concurrent GRU}) cuts the cycle count to 380, a $1045/380\approx2.75\times$ speedup over the GRU baseline (and $1201/380\approx3.16\times$ over LTC), while growing resources moderately: DSPs $44\rightarrow168$ ($\approx3.82\times$), BRAMs $7\rightarrow10$ ($\approx1.43\times$), with LUT/FF rising to $19{,}480/17{,}150$. 
Adding aggressive BRAM banking and further unrolling (\emph{BRAM-optimal GRU}) halves cycles again (380$\rightarrow$190; $2.00\times$ over Concurrent, $1045/190\approx5.50\times$ over GRU baseline, $1201/190\approx6.32\times$ over LTC), at a steep area cost: DSPs $=524$ ($11.9\times$ GRU baseline), BRAMs $=18$ ($2.57\times$), LUTs $=276{,}047$ ($26.4\times$), and FFs $=130{,}106$ ($8.38\times$).

\noindent\textbf{Throughput.}
The \emph{Interval} is the steady-state spacing (in clock cycles) between two consecutive outputs of the accelerator. 
It is the top-level initiation interval of the streaming design (under \texttt{DATAFLOW} it is essentially $\max_i \mathrm{II}_i$ of all stages, including any arbitration). 
Throughput is therefore
\[
\text{Throughput (items/s)} \;=\; \frac{F_{\max}}{\text{Interval}},
\]
$F_{\max}$ (maximum clock frequency) is the highest frequency at which a design can run \emph{without timing violations} after synthesis, placement, and routing. 
It is determined by the \emph{critical path} delay—the longest register-to-register path—plus setup/clocking overhead.
So smaller Interval $\Rightarrow$ higher throughput.


Compared to Interval, \emph{Cycles} often reports the total cycles for a given workload (e.g., to process a token/sequence or to drain the pipeline), whereas \emph{Interval} captures the steady-state rate once the pipeline is filled. 
Latency to first result depends on total pipeline depth, but long-run throughput is governed by \emph{Interval}.

The reported \emph{Interval} drops from \(12014\) (LTC) to \(271\) (GRU), \(145\) (GRU+\texttt{DATAFLOW}), and finally \(107\) (GRU+Banking). Thus, throughput—proportional to \(F_{\max}/\text{Interval}\)—improves by \(44.3\times\) when moving from LTC to GRU, by \(1.87\times\) from GRU to \texttt{DATAFLOW}, and by a further \(1.36\times\) with banking. Overall, LTC\(\rightarrow\)Banking yields \(\approx 112\times\) higher steady-state rate. The reduction from \(271\!\to\!145\!\to\!107\) aligns with (i) overlapping stages via \texttt{DATAFLOW} and (ii) adding memory ports through BRAM banking so stages can sustain \(\mathrm{II}\!\approx\!1\).




\noindent\textbf{Power and efficiency.}
From Table \ref{tab:gru-configs}, power decreases from \(5.11\,\mathrm{W}\) (LTC) \(\to\) \(4.736\,\mathrm{W}\) (GRU) \(\to\) \(3.013\,\mathrm{W}\) (Concurrent GRU), then rises to \(4.15\,\mathrm{W}\) (BRAM-banked). Assuming similar \(F_{\max}\), the energy per output \(\propto P\cdot\text{Interval}\) (normalized to LTC) is:
GRU \(=0.0209\) , which mean 97.9\% lower energy/output than LTC.
Concurrent GRU \(=0.00712\) ,
BRAM-banked \(=0.00723\).
Thus, \emph{Concurrent GRU} is the most energy-efficient by a small margin; \emph{BRAM-banked} trades a slight energy penalty for the smallest interval.


\begin{figure}[h]
\centering
\includegraphics[width=\columnwidth]{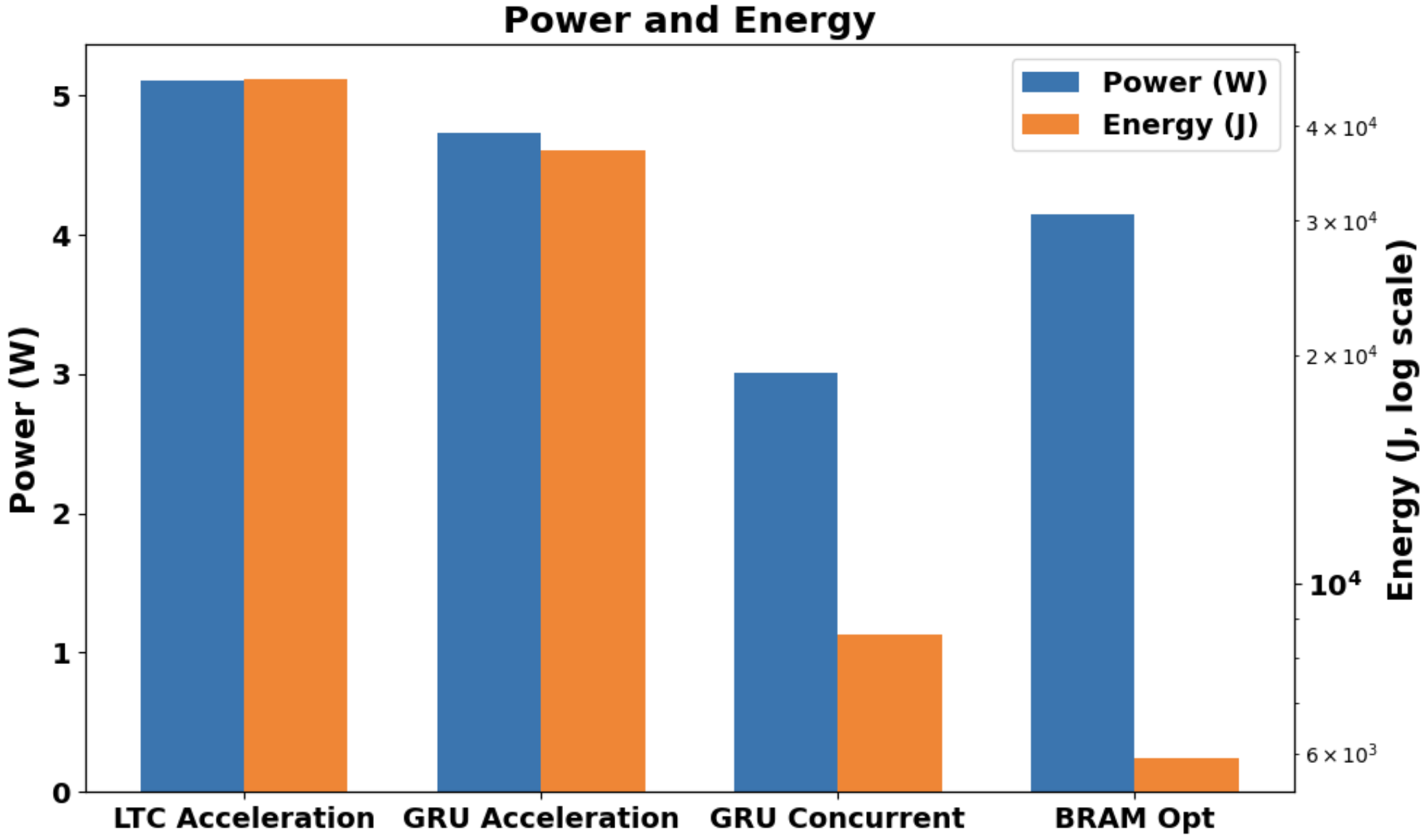}
\caption{Power (blue, linear scale) and energy (orange, log scale) consumption across four acceleration configurations: LTC Acceleration, GRU Acceleration, GRU Concurrent, and BRAM Optimization. The chart highlights how concurrent GRU mapping and BRAM optimization reduce energy significantly while keeping power moderate.}
\label{fig:power_energy_analysis}
\end{figure}




\section{Discussion}
Our experiments highlight a clear trade-off between energy/power efficiency and resource usage. 
As shown in Fig.~\ref{fig:power_energy_analysis}, the BRAM-optimal GRU design slightly increases power compared to the Concurrent GRU, but it drastically reduces total energy thanks to its much shorter execution time.

The Concurrent GRU achieves lower cycle counts and intervals with only modest increases in DSP and BRAM, making it well suited for energy- and resource-constrained settings. 
By contrast, the BRAM-optimal GRU achieves the smallest interval (highest potential throughput) but places more pressure on \(F_{\max}\), increases power, and requires significantly more LUT, FF, and DSP resources.

In practice, the preferred design depends on system priorities. 
If timing and resources are tight or energy efficiency is critical, the Concurrent GRU is preferable. 
If maximum raw throughput is the goal and sufficient area and thermal headroom are available, the BRAM-optimal GRU is the better choice because it delivers the highest parallelism and throughput.

\section{Conclusion}
We presented a hardware-conscious formulation of Model Recovery that maps naturally to FPGAs using streaming dataflow, co-designed memory/computation, and fixed-point arithmetic. This approach reduces latency and energy while maintaining accuracy, offering a practical path to real-time deployment of MR at the edge without relying on large-batch GPU execution. By aligning memory layout with compute parallelism and sustaining stage-level concurrency, the design delivers deterministic throughput and a compact DRAM footprint—properties that are especially valuable in resource- and power-constrained settings.

Looking ahead, we will systematically evaluate additional acceleration strategies from Table \ref{tab:related_work} for prior-/physics-guided modeling, including continuous-depth and ODE-targeted neural architectures. Thus far we have focused on FPGA-based accelerators; future work will explore ASIC prototypes for tighter power and area budgets, as well as analog/mixed-signal implementations. We also plan to study scale-out on multi-FPGA “towers” (clusters) and other memory-rich platforms to assess throughput and latency at larger problem sizes.

\bibliographystyle{ieeetr} 
\bibliography{mybibfile}
\end{document}